\documentclass[twocolumn,twocolappendix]{aastex631}

\usepackage{amsmath}	% Advanced maths commands
\usepackage{threeparttable}

\received{}
\revised{}
\accepted{}

\shorttitle{}
\shortauthors{}

\begin{document}

%\title{A comparison of stellar population synthesis models for epoch of reionization}
\title{A quantification of the effects using different stellar population synthesis models for epoch of reionization}
\correspondingauthor{Qingbo Ma, Yunkun Han}
\email{maqb@gznu.edu.cn, hanyk@ynao.ac.cn}

\author{Peiai Liu}
\affiliation{School of Physics and Electronic Science, Guizhou Normal University, Guiyang 550001, PR China}

\author[0000-0001-9493-4565]{Qingbo Ma}
\affil{School of Physics and Electronic Science, Guizhou Normal University, Guiyang 550001, PR China}
\affil{Guizhou Provincial Key Laboratory of Radio Astronomy and Data Processing, \\
Guizhou Normal University, Guiyang 550001, PR China}

\author{Yunkun Han}
\affil{Yunnan Observatories, Chinese Academy of Sciences, 396 Yangfangwang, Guandu District, Kunming, 650216, PR China}

\author{Rongxin Luo}
\affil{School of Physics and Electronic Science, Guizhou Normal University, Guiyang 550001, PR China}
\affil{Guizhou Provincial Key Laboratory of Radio Astronomy and Data Processing, \\
Guizhou Normal University, Guiyang 550001, PR China}

\begin{abstract}
The luminosity and spectral energy distribution (SED) of high-$z$ galaxies are sensitive to the stellar population synthesis (SPS) models.
In this paper, we study the effects of different SPS models on the measurements of high-$z$ galaxies and the budget of ionizing photons during the epoch of reionization, by employing each of them in the semi-analytical galaxy formation model {\sc L-Galaxies 2020}.
We find that the different SPS models lead to $\lesssim 0.5$ dex differences on the amplitudes of UV luminosity functions, while the two modes of the same SPS model with and without the inclusion of binary stars leads to similar UV luminosity functions at $z \ge 6$.
Instead, the binary stars produce $\sim 40\%$ more ionizing photons than the single stars, while such differences are smaller than those caused by different SPS models, e.g. the BPASS model produces $\sim 100\%$ more ionizing photons than other models.
\end{abstract}


\section{Introduction}
Although the gas of intergalactic medium (IGM) is observed highly ionised in today's Universe, it was fully neutral in the past after the cosmic recombination, the period of transition from neutral to ionised  phase of the Universe is called as the epoch of reionization \cite[EoR, ][]{Furlanetto2006, Dayal2018}.
EoR happened $\sim$ 1.3 million years after the big bang, when the first generation of stars and galaxies formed and radiated photons, which then ionised the hydrogen and helium in the IGM \citep{Barkana2016}.
The observations, e.g. Ly$\alpha$ absorption lines of quasars \citep{Fan2006}, the optical depth of cosmic microwave background radiation \cite[CMB, ][]{Planck2020} and the Ly$\alpha$ emitters \citep{Weinberger2019}, suggest that the EoR ended at $z > 5$.

The evolution of EoR includes the physical processes of the formation of dark matter halos and large-scale structures, the formation of first stars and galaxies, the radiation of UV and X-ray sources, and the ionizing and heating of the IGM gas \citep{Furlanetto2006}.
The galaxy formation relates to e.g. the cooling of hot gas, star formation, supernova feedback, active galactic nuclei (AGN) feedback and galaxy mergers \citep{Dayal2018}.
Some theoretical models have been developed to describe the galaxy formation and evolution, e.g. the halo occupation distribution model \cite[HOD, ][]{Zheng2005}, the sub-halo abundance matching model \cite[SHAM, ][]{Campbell2018} and the conditional luminosity function model \cite[CLF, ][]{Yang2003}.
These models focus on the correlations between different physical quantities on the formation and evolution of galaxies, while do not involve specific physical processes.
The hydrodynamic simulations can include abundant physical processes and provide the best description of galaxy formation processes \citep{Kannan2022}, while they are very computing expensive.
The semi-analytical model (SAM) basing on the merger trees from {\it N}-body simulations can describe almost all the physical processes related to galaxy formation \citep{Henriques2020}, which is more efficient than the hydrodynamic simulations but more precise than the theoretical models.
The SAM models, e.g. the Santa Cruz semi-analytic model \citep{Yung2019}, {\sc ASTRAEUS} \citep{Hutter2021} and {\sc MERAXES} \citep{Balu2023}, have been applied to study the high-$z$ galaxies ($z \ge 6$), which can explain the high-$z$ measurements e.g. the UV luminosity functions.
Recently, the well-developed galaxy formation SAM model {\sc L-Galaxies 2020} \citep{Henriques2020} is also applied to explain the high-$z$ observations and study the EoR process \citep{Ma2023}.

The stellar population synthesis (SPS) model is a key component within galaxy formation models \citep{Conroy2013}, which relates the stellar mass, age, and metallicity to the luminosity and spectral energy distribution (SED) of galaxies \citep{Henriques2015}.
Many SPS models have been developed to explain the observations of galaxies, e.g. \citealt{Bruzual2003} (named as BC03), \citealt{Maraston2005} (named as M05), Yunnan evolutionary population synthesis model \cite[YEPS, ][]{zhang2004, Zhang2005}, and Binary Population and Spectral Synthesis mode \cite[BPASS, ][]{Eldridge2017, Stanway2018}.
Since more than 50\% of observed stars in galaxies and clusters are in binary systems, the YEPS and BPASS models also consider the effects of binary stars \citep{Zhang2005, Stanway2018}.
The interaction of binary stars can change the expected SEDs of galaxies and produce more ionising photons \citep{Stanway2016, Gotberg2020}.
The differences on the SPS models are expected to affect the properties of high-$z$ galaxies and the budget of ionizing photons \citep{Wilkins2016, Seeyave2023}, which can be measured by the current and near future telescopes.
For example, the high-$z$ galaxies can be observed by the Hubble Space Telescope \cite[HST, ][]{Bouwens2015} and the James Webb Space Telescope \cite[JWST, ][]{Roberts-Borsani2021}, while the EoR can be measured by the 21-cm signals from neutral hydrogen with the low-frequency radio telescope arrays such as the Low Frequency Array (LOFAR\footnote{\url{http://www.lofar.org/}}), the Square Kilometre Array (SKA\footnote{\url{https://www.skatelescope.org/}}), the Murchison Widefield Array (MWA\footnote{\url{http://www.mwatelescope.org/}}), and the Hydrogen Epoch of Reionization Array (HERA\footnote{\url{https://reionization.org/}}).

In this paper, we investigate the effects of different SPS models on the observations of high-$z$ galaxies and the budget of ionizing photons using the SAM model {\sc L-Galaxies 2020} \citep{Henriques2020} and the Millennium-II simulation \citep{Boylan-Kolchin2009}.
The paper is organized as the following: we describe the galaxy formation and SPS models adopted in Sect \ref{sec:method}, present the results in Sect \ref{sec:res}, and the conclusions are summarized in Sect \ref{sec:conclusion}.

%%%%%%%%%%%%%%%%%%%%%%%%%%%%%%%%%%%%%%%%%%%%%%%%%%
\section{Methods}
\label{sec:method}

We apply the SAM model {\sc L-Galaxies 2020} \citep{Henriques2020} in combination with the {\it N}-body dark matter simulation Millennium-II \citep{Boylan-Kolchin2009} to investigate the effects of different SPS models on the study of high-$z$ galaxies and EoR.
We will briefly describe the simulations and the SPS models here, while the readers can refer to the original papers for more details.

%%%%%%%%%%%%%%%%%%%%%%%%%%%%%%%%%%%%%%%%%%%%%%%%%%
\subsection{Dark matter simulations}

The dark matter simulation and merger trees adopted are from the Millennium-II simulation \citep[MS-II, ][]{Boylan-Kolchin2009}, which was run with an updated version of the {\sc GADGET} code \citep{Springel2005} i.e. {\sc GADGET-3}.
The original box size of MS-II is $100\, {\rm Mpc}/h$ with $2160^{3}$ dark matter particles, and each particle is with mass of $6.89 \times10^{6} \, h^{-1}\, {\rm M_\odot}$.
The halos were identified using the Friends-Of-Friends (FOF) algorithm \citep{More2011}, then the {\sc SUBFIND} algorithm was applied to identify the self-bound substructures within each FOF group.
The halos are with at least 20 particles, i.e. the minimal halo mass $1.38 \times10^{8} \, h^{-1}\, {\rm M_\odot}$.
The simulations are scaled to the {\it Planck} cosmology \citep{Planck2020} following the procedure of \cite{Angulo2015}, with the cosmological parameters $\Omega_{m} = 0.315$, $\Omega_{b} = 0.049$, $\Omega_{\Lambda}= 0.685$,  $h = 0.673$, $\sigma_{8} = 0.826$ and $n_{s} = 0.965$.

MS-II simulation has outputs of 68 snapshots from $z=127$ to 0, while 22 of them are at $z \ge 6$.
The merger trees were constructed with the sub-halos found in these snapshots, in this case the merger trees of MS-II simulation include $\sim$ 590 million sub-halos in total \citep{Boylan-Kolchin2009}.
The more or less snapshots should not obviously change the conclusions presented in this paper.
We test it with the low resolution simulations described in \cite{Ma2023}.

%%%%%%%%%%%%%%%%%%%%%%%%%%%%%%%%%%%%%%%%%%%%%%%%%%
\subsection{SAM model {\sc L-Galaxies 2020}}

We adopt the public SAM model {\sc L-Galaxies 2020} \citep{Henriques2020} to evolve the galaxies with the merger trees from MS-II simulation.
This model has included almost all the physical processes related to galaxy formation, such as gas cooling, star formation, galaxy merger, supernovae and AGN feedback.
The application of {\sc L-Galaxies 2020} on the high-$z$ galaxies and EoR has been explored in \cite{Ma2023}.
The {\sc L-Galaxies 2020} model is the updated version of \cite{Henriques2015}, which has a few differences with respect to the latter.
In the new model, the galactic discs are spatially resolved by dividing the discs into $12$ concentric rings with radii $r_i = 0.01 \times 2^i h^{-1} \, {\rm kpc},\,i=0,...,11$ \citep{Fu2013}.
All the properties and physical processes of discs, such as star formation, chemical enrichment and gas ejection, are evolved for each ring separately.
The star formation is linked to the H$_2$ abundance of each ring that depending on the metallicity of gas \citep{Krumholz2009, McKee2010}.

We adopt the default values for the free parameters in the {\sc L-Galaxies 2020} code, which are the best fit values with the observations of galaxies at low-$z$ and the merger trees from MS-II simulation \citep{Henriques2020}.
We only modify the input SPS model adopted in the {\sc L-Galaxies 2020} code.

%%%%%%%%%%%%%%%%%%%%%%%%%%%%%%%%%%%%%%%%%%%%%%%%%%
\subsection{Stellar populations synthesis}
\begin{table*}
    \begin{threeparttable}
    \begin{center}
    \caption{Parameters of four SPS models applied in the {\sc L-Galaxies 2020} model.}
    \label{tab:model_input}
    \begin{tabular}{lllllll}
        \hline
        Models &  Wavelength range    & $N_{\text{wavelength}}$   & Age range      & $N_{\text{ages}}$ &$N_{\text{metallicites}}$  \\
        \hline
        BC03   & 91 \AA-$3.6 \times 10^8$ \AA     & 2023            & 0 \,yr\,-\,20 \,Gyr      & 221                    & 7  \\
        M05\tnote{a}    & 91 \AA-$1.6 \times 10^6$ \AA     & 1221                & $10^3$\, yr\,-\,15 \,Gyr  & 67                     & 4  \\
        % CB07   & Salpeter  & 91 \AA-$3.6 \times 10^8$ \AA     & 2023                     & 0 Myr-20 Gyr      & 221                    & 7  \\
        YEPS\tnote{b}   & 91 \AA-$1.6 \times 10^6$ \AA     & 1221            &0.1 \,Myr\,-\,15 \,Gyr     & 90                     & 7  \\
        BPASS  &  1\, \AA-$1.0 \times 10^5$ \AA     &100,000              &1 \,Myr\,-\,100\, Gyr      & 51                     & 13 \\
        \hline
    \end{tabular}
    \begin{tablenotes}
    \footnotesize
    \item[a] The version incorporating the red horizontal-branch morphology was employed.
    \item[b] The version employed in this paper is Yunnan-II, while the updated Yunnan-III version \citep{yepsiii2013} is only available for solar-metallicity.
    \end{tablenotes}
    \end{center}
    \end{threeparttable}
\end{table*}
We adopt four popular SPS models in this paper, i.e. BC03 \citep{Bruzual2003}, M05 \citep{Maraston2005}, YEPS \citep{Zhang2005} and BPASS \citep{Stanway2018}.
The parameters of these SPS models, including the wavelength range, age and metallicity, are summarized in Tab.~\ref{tab:model_input}.
To make consistent comparison, we employ the results of these models with the same Salpeter stellar initial mass function \cite[IMF,][]{Salpeter1955} with a power index $\alpha$ of 1.35 and maximum initial mass 100 $\rm M_\odot$.
For the YEPS and BPASS models, we consider both the ones including binary stars, which are named as YEPS~BS and BPASS~BS respectively, and the ones with only single stars, which are named as YEPS~SS and BPASS~SS respectively.

\begin{figure*}
\centering
    \includegraphics[width=0.95\linewidth]{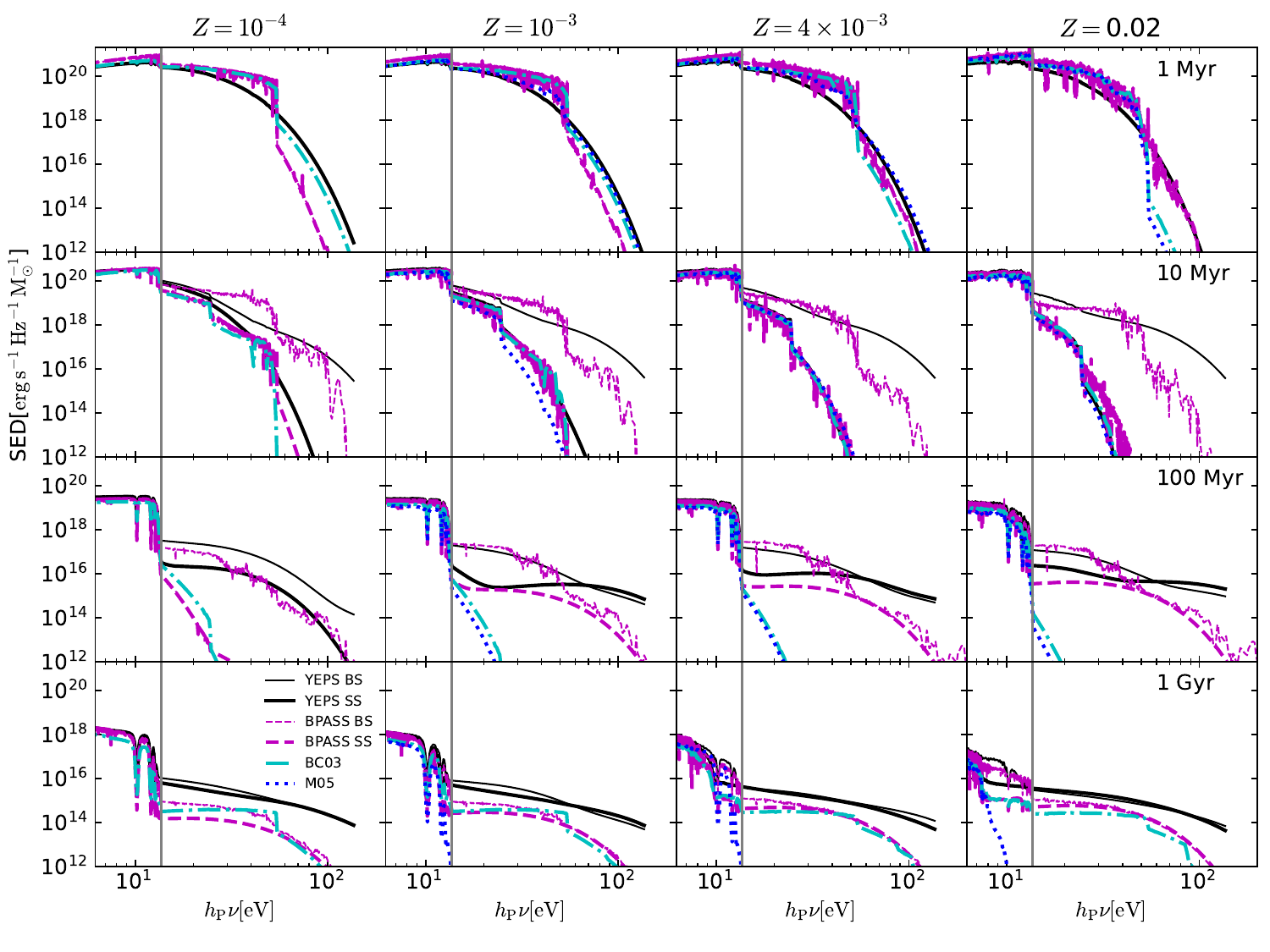}
    \caption{Stellar mass normalized SEDs of four SPS models, i.e. BC03 (dash-dotted cyan), M05 (dotted blue), YEPS (solid black) and BPASS (dashed magenta), at metallicity $Z=10^{-4}$, $10^{-3}$, $4 \times 10^{-3}$ and 0.02 (from left to right), and age 1 Myr, 10 Myr, 100 Myr and 1 Gyr (from top to bottom).
    The thin lines of YEPS (YEPS~BS) and BPASS (BPASS~BS) models are the SEDs including binary stars, while the thick lines (i.e. YEPS~SS and BPASS~SS) are with single stars.
    Note that in the BC03 model, the SED output of $Z=10^{-3}$ is obtained by interpolation.
    In the M05 model, the SED output of $Z=4 \times 10^{-3}$ is obtained by interpolation, and at $Z=10^{-4}$ its SED is not shown because the age of star is only in the range of 1 Gyr-15 Gyr.
    The vertical gray lines denote the location of $13.6\,\rm eV$.}
    \label{fig:origizal_sed_inputs}
\end{figure*}
As a reference, in Fig.~\ref{fig:origizal_sed_inputs} we present the initial SEDs from different SPS models at four metallicities ($Z=10^{-4}$, $10^{-3}$, $4 \times 10^{-3}$ and 0.02) and four ages (1 Myr, 10 Myr, 100 Myr and 1 Gyr).
Note that the SEDs in Fig.~\ref{fig:origizal_sed_inputs} are only in the frequency range $[6.2, 200]\, \rm eV$ (i.e. UV band), while the Fig.~\ref{fig:origizal_sed_inputs_full} in Appendix~\ref{app:full_sed} shows one example of the full SEDs.
As showed in Fig.~\ref{fig:origizal_sed_inputs}, the initial SEDs from all four SPS models are not very sensitive to the metallicities, while the amplitudes reduce obviously with the increasing ages.
The different SPS models present roughly consistent SEDs, especially at $h_{\rm P}\nu <13.6\,\rm eV$, while some differences are still visible.
The inclusion of binary stars in the YEPS and BPASS models (i.e. YEPS~BS and BPASS~BS) leads to harder SEDs at $h_{\rm P}\nu > 13.6\,\rm eV$ compared to those with single stars (i.e. YEPS~SS and BPASS~SS), and thus expects to increase the emission of ionizing photons.
In the following, we will briefly describe these SPS models.
For more comparisons about the SPS models, one can also refer to e.g. \cite{Chen2010} and \cite{Han2019}.

%%%%%%%%%%%%%%%%%%%%%%%%%%%%%%%%%%%%%%%%%%%%%%%%%%
\subsubsection{BC03 Model}

The SED of BC03\footnote{\url{http://www.bruzual.org/}} covers the wavelength range from 91 \AA~to $3.6 \times 10^{8}$ \AA~with 2023 outputs.
The age of stars is from 0~yr to 20~Gyr, with a grid of 221 steps.
There are seven available metallicities, i.e. $Z=10^{-4}$, $4 \times 10^{-4}$, $4 \times 10^{-3}$, $8 \times 10^{-3}$, 0.02, 0.05 and 0.1.
This information is summarized in Tab.~\ref{tab:model_input}.
From Fig.~\ref{fig:origizal_sed_inputs}, the stellar mass normalized SEDs from BC03 are similar to the M05 model at age $\le$ 100 Myr, while closer to the BPASS model at 1 Gyr.
With the increasing of metallicities, the radiations at $h_{\rm P}\nu>13.6\,\rm eV$ are slightly reduced at age 10 Myr, while this effect is not very significant at other ages.

%%%%%%%%%%%%%%%%%%%%%%%%%%%%%%%%%%%%%%%%%%%%%%%%%%
\subsubsection{M05 Model}

The wavelength range of the SED from M05\footnote{\url{http://www.icg.port.ac.uk/~maraston/}} is from 91 \AA~to $1.6\times 10^{6}$ \AA, with 1221 outputs.
The age of stars is $10^3$~yr to 15~Gyr, with a grid of 67 steps.
The outputs are available at four metallicity values, i.e. $Z=10^{-3}$, 0.01, 0.02 and 0.04.
As showed in Fig.~\ref{fig:origizal_sed_inputs}, the SEDs of M05 are roughly similar to the BC03 model at age $\le$ 100 Myr, while present obviously less high energy radiation ($h_{\rm p}\nu > 13.6\,\rm eV$) at 1 Gyr.
Its SEDs are also similar to the single star mode of YEPS and BPASS models (i.e. YEPS~SS and BPASS~SS) at age 1 Myr and 10 Myr, while have less radiation at $h_{\rm p}\nu > 13.6\,\rm eV$ at age 100 Myr and 1 Gyr.

%%%%%%%%%%%%%%%%%%%%%%%%%%%%%%%%%%%%%%%%%%%%%%%%%%
\subsubsection{YEPS model}

The wavelength range of YEPS SED\footnote{\url{http://www1.ynao.ac.cn/~zhangfh/YN_SP.html}} is 91~\AA~- $1.6\times 10^{6}$ \AA~with 1221 outputs.
The age of stars covers the range from 0.1 Myr to 15 Gyr, with 90 outputs.
The results at 7 metallicities are available, i.e. $Z=10^{-4}$, $3 \times 10^{-4}$, $10^{-3}$, $4 \times 10^{-3}$, 0.01, 0.02, 0.03.
We adopt two modes of SEDs from YEPS model, one with the binary stars (YEPS~BS) and one with only the single stars (YEPS~SS).
From Fig.~\ref{fig:origizal_sed_inputs}, the SEDs of single star mode (YEPS~SS) are similar to other models at 1 Myr and 10 Myr, while close to the BPASS model at 100 Myr and 1 Gyr.
Compared to YEPS~SS, the differences caused by the binary stars (YEPS~BS) are significant at age 10 Myr and slightly at 100 Myr, but not too much at 1 Myr and 1 Gyr.
The inclusion of binary stars obviously increases the amplitudes of SEDs at $h_{\rm p}\nu > 13.6\,\rm eV$ at age 10 Myr, while not too much at other ages.

%%%%%%%%%%%%%%%%%%%%%%%%%%%%%%%%%%%%%%%%%%%%%%%%%%
\subsubsection{BPASS model}

We adopt the results of version 2.2.1 of BPASS\footnote{\url{http://bpass.auckland.ac.nz}} model, including the modes with binary stars (BPASS~BS) and only single stars (BPASS~SS).
The wavelength range is from 1~\AA~to $10^{5}$~\AA~with $10^{5}$ outputs.
The age of stars is from 1 Myr to 100 Gyr with 51 steps.
There are 13 available metallicities: $Z=10^{-5}$, $10^{-4}$, $10^{-3}$, $2 \times 10^{-3}$, $3 \times 10^{-3}$, $4 \times 10^{-3}$, $6 \times 10^{-3}$, $8 \times 10^{-3}$, 0.01, 0.014, 0.02, 0.03, 0.04.
As showed in Fig.~\ref{fig:origizal_sed_inputs}, the SEDs of BPASS model have slightly higher amplitudes than other models at $h_{\rm p}\nu < 50\,\rm eV$ and age 1 Myr.
At 10 Myr, its SEDs with single stars (BPASS~SS) are similar to other models.
At 100 Myr and 1 Gyr, the SEDs of BPASS~SS are slightly lower than the YEPS model, but have more high energy radiation ($h_{\rm p}\nu > 13.6\,\rm eV$) than the BC03 and M05 models.
The inclusion of binary stars (BPASS~BS) increases the SEDs at $h_{\rm p}\nu > 13.6\,\rm eV$ at age 10 Myr, but not too much at other ages, similar to that of YEPS model.

%%%%%%%%%%%%%%%%%%%%%%%%%%%%%%%%%%%%%%%%%%%%%%%%%%%%%%%%%%%%%
\section{Results}
\label{sec:res}

In this section, we present the properties of high-$z$ galaxies and the budgets of ionizing photon from {\sc L-Galaxies 2020} with different SPS models.

%%%%%%%%%%%%%%%%%%%%%%%%%%%%%%%%%%%%%%%%%%%%%%%%%%
\subsection{SEDs of high-$z$ galaxies}

\begin{figure*}
\centering
    \includegraphics[width=0.95\linewidth]{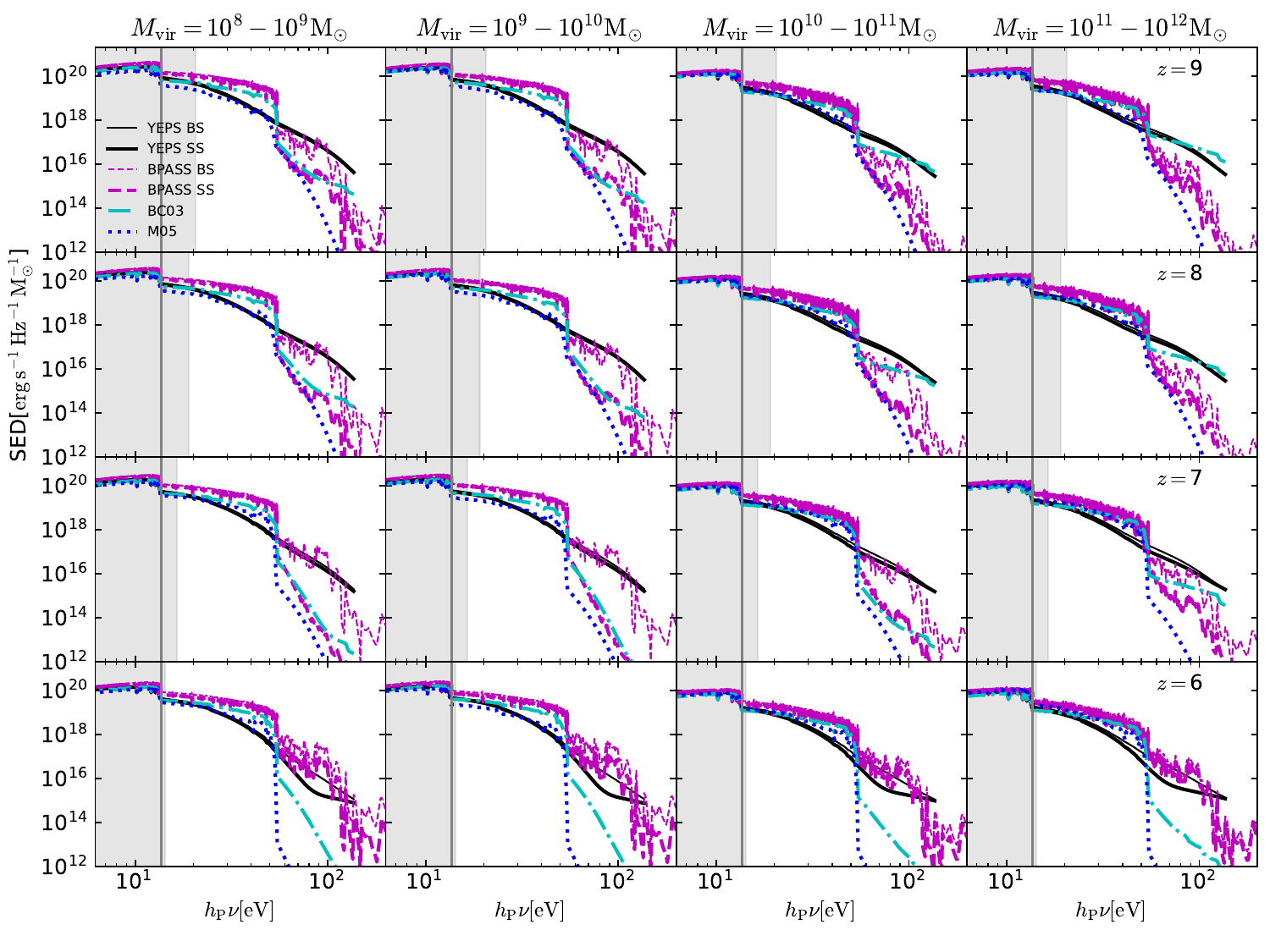}
    \caption{Stellar mass normalized SEDs of high-$z$ galaxies with halo mass (from left to right) $M_{\rm vir} = 10^{8} - 10^{9}\,\rm M_{\odot}$, $10^{9} - 10^{10}\,\rm M_{\odot}$, $10^{10} - 10^{11}\,\rm M_{\odot}$ and $10^{11} - 10^{12}\,\rm M_{\odot}$ with four SPS models i.e. BC03 (dash-dotted cyan),  M05 (dotted blue), YEPS (solid black) and BPASS (dashed magenta).
    From top to bottom, the results are at $z=9$, 8, 7 and 6.
    The thin lines of YEPS (YEPS~BS) and BPASS (BPASS~BS) models are the SEDs including binary stars, while the thick lines (i.e. YEPS~SS and BPASS~SS) are with single stars.
    The vertical gray lines denote the location of $13.6\,\rm eV$.
    The semitransparent gray region denotes the frequency ranges that can be measured by the JWST telescope.
    }
    \label{fig:output_sed_multi_z}
\end{figure*}

The SEDs of galaxies from {\sc L-Galaxies 2020} are computed with the SPS results (i.e. the Fig.~\ref{fig:origizal_sed_inputs}) and the history of star formation and metal enrichment within each galaxy.
More specifically, with the information of star formation and metal enrichment history produced, the {\sc L-Galaxies 2020} post-processes the SEDs of each galaxy by linearly interpolated the input SEDs from different SPS models at specific ages and metallicities, and then multiply the star mass.
The final SED of galaxies is the sum of the results at all ages.
Fig.~\ref{fig:output_sed_multi_z} shows the average rest-frame SEDs of high-$z$ galaxies within the same halo mass range obtained from {\sc L-Galaxies 2020} with four SPS models at different $z$s.
To compare with the input SEDs showed in Fig.~\ref{fig:origizal_sed_inputs}, the results of Fig.~\ref{fig:output_sed_multi_z} are normalized by the stellar mass of galaxies.
By comparing with the Fig.~\ref{fig:origizal_sed_inputs}, we can see that the SEDs of high-$z$ galaxies within UV band are dominated by the young stars, i.e. those with age $<$ 100 Myr.
Due to the same reason, the SEDs of high-$z$ galaxies do not evolve too much with the decreasing redshift.
The SEDs of massive halos are slightly lower than the less massive ones, since the latter ones have higher ratio of star formation rate over stellar mass \citep{Henriques2020}.

Four SPS models predict similar SEDs of high-$z$ galaxies at $h_{\rm p}\nu <13.6 \,\rm eV$, while some differences are obvious at higher energy band.
The BPASS model shows obviously higher amplitudes of SEDs than other models at $h_{\rm p}\nu = 13.6 - 50 \,\rm eV$.
Within the same band, the BC03 model is slightly lower than the BPASS model, while the YEPS model has the lowest amplitude, and the M05 model is between BC03 and YEPS models.
Differently, the YEPS model has the highest SED of galaxies at $h_{\rm p}\nu > 50 \,\rm eV$ and $z\ge 7$, while becomes similar to the BPASS model at $z=6$.
Within the same band, the SEDs of galaxies with M05 model are roughly consistent with the BPASS model at $z\ge 7$, while obviously lower than the latter at $z=6$.
The SEDs of galaxies with BC03 model at $z\ge 7$ are similar to those with M05 and BPASS models at $M_{\rm vir} < 10^{10}\,\rm M_{\odot}$, while become similar to those with YEPS models with the increasing halo mass e.g. at $M_{\rm vir} = 10^{11} - 10^{12}\,\rm M_{\odot}$.
The inclusion of binary stars (BPASS~BS) does not obviously change the SEDs of BPASS model at $h_{\rm p}\nu < 50 \,\rm eV$, while increases the high energy radiation at $h_{\rm p}\nu > 50 \,\rm eV$.
Such effect is weaker on the SEDs of massive halos (e.g. $M_{\rm vir} = 10^{11} - 10^{12}\,\rm M_{\odot}$) than the less massive ones, due to the increasing contributions of old stars (age $>$ 100 Myr).
With the same reason, the SEDs of galaxies at $z=6$ has no significant features of binary stars.
Instead, the inclusion of binary stars in YEPS model (YEPS~BS) has no obvious effects on the SEDs of high-$z$ galaxies, except slightly higher amplitude at $h_{\rm p}\nu > 40 \,\rm eV$ and $z=6$.

Note that, due to the abundant neutral hydrogen during EoR, it will be hard to measure the SEDs at $> 13.6\, \rm eV$, although the uncertainties of different SPS models are mostly at such band.
The measurements and comparisons at rest-frame UV band e.g. the UV luminosity function ($\phi$ showed in Fig.~\ref{fig:uvlf_multi_z}) can exclude the uncertainties of physical processes except the SPS models, while the results of SED fitting, e.g. the ionizing photon production efficiency $\zeta_{\rm ion}$, can help to distinguish the different SPS models \citep{Seeyave2023}.

%%%%%%%%%%%%%%%%%%%%%%%%%%%%%%%%%%%%%%%%%%%%%%%%%%
\subsection{UV luminosity function of galaxies}
\begin{figure}
\centering
    \includegraphics[width=0.98\linewidth]{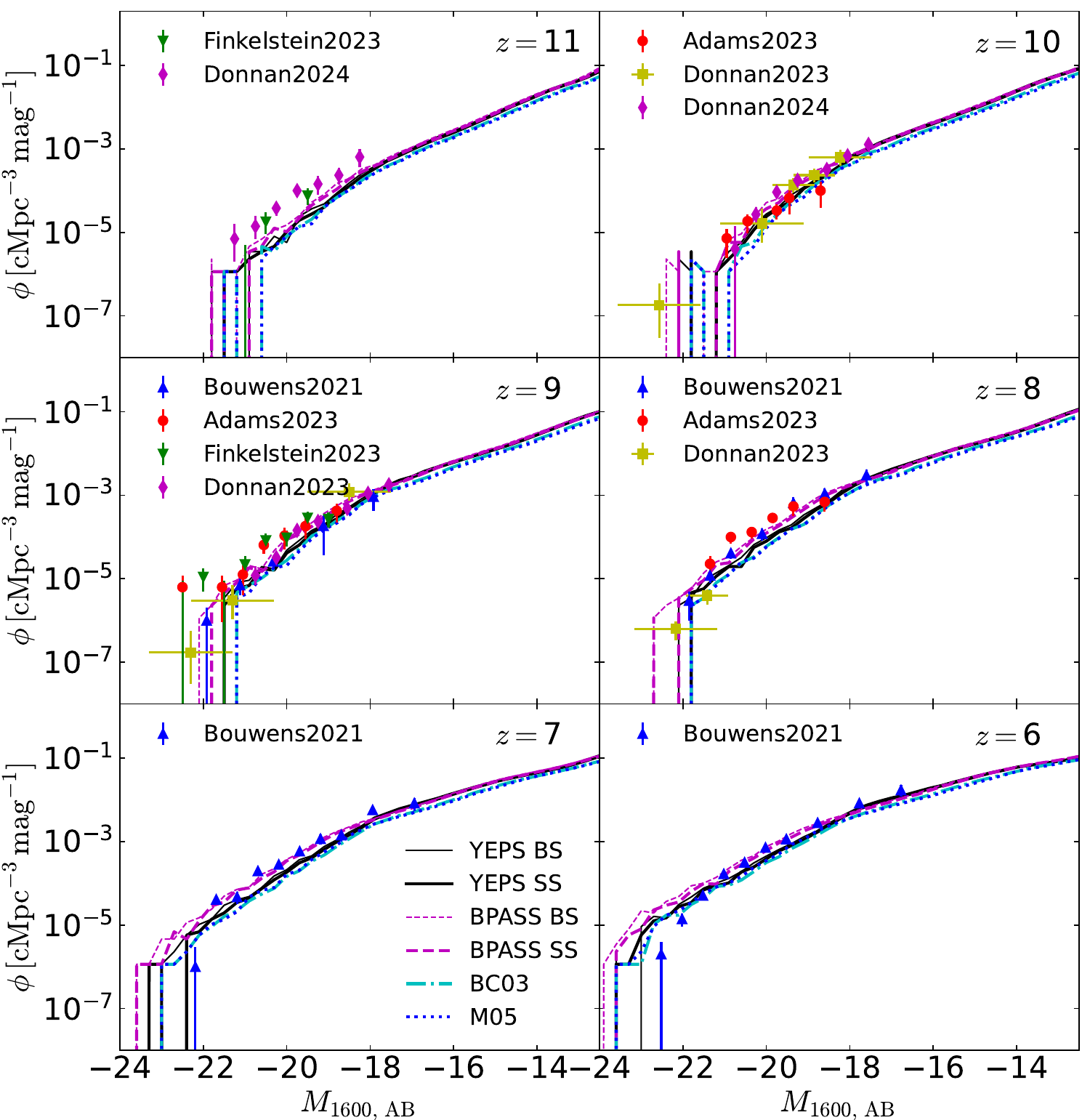}
    \caption{UV luminosity function $\phi$ at the rest-frame wavelength $\lambda = 1600$ \AA~from four SPS models i.e. BC03 (dash-dotted cyan), M05 (dotted blue),  YEPS (solid black) and BPASS (dashed magenta).
    The thin lines of YEPS (YEPS~BS) and BPASS (BPASS~BS) models are the results including binary stars, while the thick lines (i.e. YEPS~SS and BPASS~SS) are with single stars.
    From left to right and top to bottom, the six panels are the results at $z = 11$, 10, 9, 8, 7 and 6, respectively.
    The observational data points are from \citealt{Bouwens2021} (blue up triangle), \citealt{Adams2023} (red circle), \citealt{Finkelstein2023} (green down triangle), \citealt{Donnan2023} (yellow square) and \citealt{Donnan2024}  (magenta diamond).
    Note that the observations of \citealt{Bouwens2021} are at $\lambda = 1600$ \AA, while others are at $\lambda = 1500$ \AA.}
    \label{fig:uvlf_multi_z}
\end{figure}
Fig.~\ref{fig:uvlf_multi_z} shows the UV luminosity function $\phi$ at the rest-frame wavelength $\lambda = 1600$ \AA~from four SPS models at different $z$s.
The absolute magnitude of galaxy luminosity at $\lambda = 1600$ \AA~is computed by:
\begin{equation}
    M_{1600,\rm AB} = -\frac{5}{2}{\rm log}_{10}\left(\frac{F_{1600}}{4\pi R^{2}}\right) -48.6
\end{equation}
where $F_{1600}$ is the brightness of galaxy at $\lambda = 1600$\AA, and $R=10\,\rm pc$.
As a comparison, we also present some of recent observations of $\phi$ from HST \citep{Bouwens2021} and JWST \citep{Donnan2023, Adams2023, Finkelstein2023, Donnan2024} telescopes.
Note that the results of \cite{Bouwens2021} are at $\lambda = 1600$\AA, while others that observed by the JWST telescope are at $\lambda = 1500$ \AA.
Such differences are not significant on the UV luminosity functions.
Table~\ref{tab:chis} shows the $\chi^2$ of $\phi$ from different SPS models compared to the observations at three $z$s, which is defined as:
\begin{equation}
    \chi^2 = \sum \frac{(\phi_{\rm obs} - \phi_{\rm sim})^2}{\sigma_{\rm obs}^2}
\end{equation}
where $\phi_{\rm obs}$ is the observational $\phi$, $\phi_{\rm sim}$ is the $\phi$ from simulations, and $\sigma_{\rm obs}$ is the 1-$\sigma$ error of $\phi_{\rm obs}$.
Due to the lack of bright galaxies from MS-II simulation, the $\chi^2$ is computed only for the galaxies with $M_{1600,\rm AB}>-21$.

\begin{table*}
    \caption{$\chi^2$ of UV luminosity function $\phi$ from four SPS models compared to the observations at $z=10$, 9 and 8.}
    \begin{threeparttable}
    \label{tab:chis}
    \begin{tabular}{lcccccc}
        \hline
        $z$ &  \, YEPS BS \, & \, YEPS SS \, & \, BPASS BS \, & \, BPASS SS \, & \, BC03 \, & \, M05  \\
        \hline
        10 &  16.49 & 20.26 & 26.39 & 18.10 &  28.81 & 32.63 \\
        9 & 22.02 & 29.01 & 26.02 & 14.62 & 40.75 & 45.30 \\
        8 & 25.75 & 31.80 & 11.83 & 16.27 & 43.03 & 43.99 \\
        \hline
    \end{tabular}
    \end{threeparttable}
\end{table*}

The UV luminosities of galaxies can be reduced by the extinction models of dust and molecular clouds.
We test the dust model within {\sc L-Galaxies 2020}, which does not change too much on the Fig.~\ref{fig:uvlf_multi_z}, especially at $z>7$.
This is different to previous studies, e.g. the results of {\sc L-Galaxies 2015} \citep{Clay2015}.
Although the dust model is not included to precisely match the results from the MS-II simulation and {\sc L-Galaxies 2020} code with the observations, it indeed affects the UV luminosity of the bright galaxies at low redshifts \citep{Yung2020b, Bhagwat2023}, but not too much on the ones at high redshifts and the faint ones.

The UV luminosity function $\phi$ from four SPS models are roughly consistent with the observations at six $z$s.
The differences of four SPS models on $\phi$ are $\lesssim 0.5$ dex, smaller than the uncertainties (i.e. error bars) of the current measurements of $\phi$.
Specifically, the BPASS model has higher amplitudes of $\phi$ than other models.
The BC03 model is globally similar to the M05 model, which $\phi$s have amplitudes lower than the YEPS and BPASS models.
The YEPS model is similar to the BPASS model at $M_{1600,\rm AB}>-18$, while closer to the BC03 and M05 models at $M_{1600,\rm AB}<-18$.
The inclusion of binary stars has negligible effects on $\phi$, both for the YEPS and BPASS models, consistent with the SEDs of high-$z$ galaxies showed in Fig.~\ref{fig:output_sed_multi_z}.
With the $\chi^2$ showed in Table~\ref{tab:chis}, at $z=10$ the YEPS BS model fits better with the observations, then is the BPASS SS models. At $z=9$, the BPASS SS model is the best fit one. At $z=8$, the best fit model is the BPASS BS model, then is the BPASS SS model.

%%%%%%%%%%%%%%%%%%%%%%%%%%%%%%%%%%%%%%%%%%%%%%%%%%
\subsection{Budget of ionizing photons}

To properly compute the number of ionizing photons from high-$z$ galaxies, we rerun the {\sc L-Galaxies 2020} SAM simulations with the integrated SED (iSED) over the age of stars (i.e. the time from the birth of stars to the output $z$s) for four SPS models.
The iSED can easily include the effects of star formation history \citep{Ma2023}.
Meanwhile, as showed in \cite{Ma2023}, after normalized by the stellar mass the iSED of high-$z$ galaxies is sensitive neither to the galaxy formation models nor to the output $z$s.
The number of ionizing photons ($n_{\rm ion}$) from galaxies is then calculated by:
\begin{equation}
n_{\rm ion} = \int_{\rm 13.6\,eV} \frac{L_{\nu}}{h_{\rm P} \nu} \mathrm{d}\nu
\end{equation}
where $L_{\nu}$ is the iSED of galaxies computed by {\sc L-Galaxies 2020}, $h_{\rm P}$ is the Planck constant, and $\nu$ is the frequency of photons.
The integration is done with the full iSED at $h_{\rm p}\nu > 13.6 \,\rm eV$.
The cosmic volume averaged ionizing photons $N_{\rm ion}$ (i.e. number density of ionizing photon) is expressed as:
\begin{equation}
N_{\rm ion} = \frac{\sum n_{\rm ion}}{V_{\rm box}}
\end{equation}
where the sum $\sum$ is for all the selected galaxies, and $V_{\rm box}$ is the comoving volume of MS-II simulation.

\begin{figure}
\centering
    \includegraphics[width=0.98\linewidth]{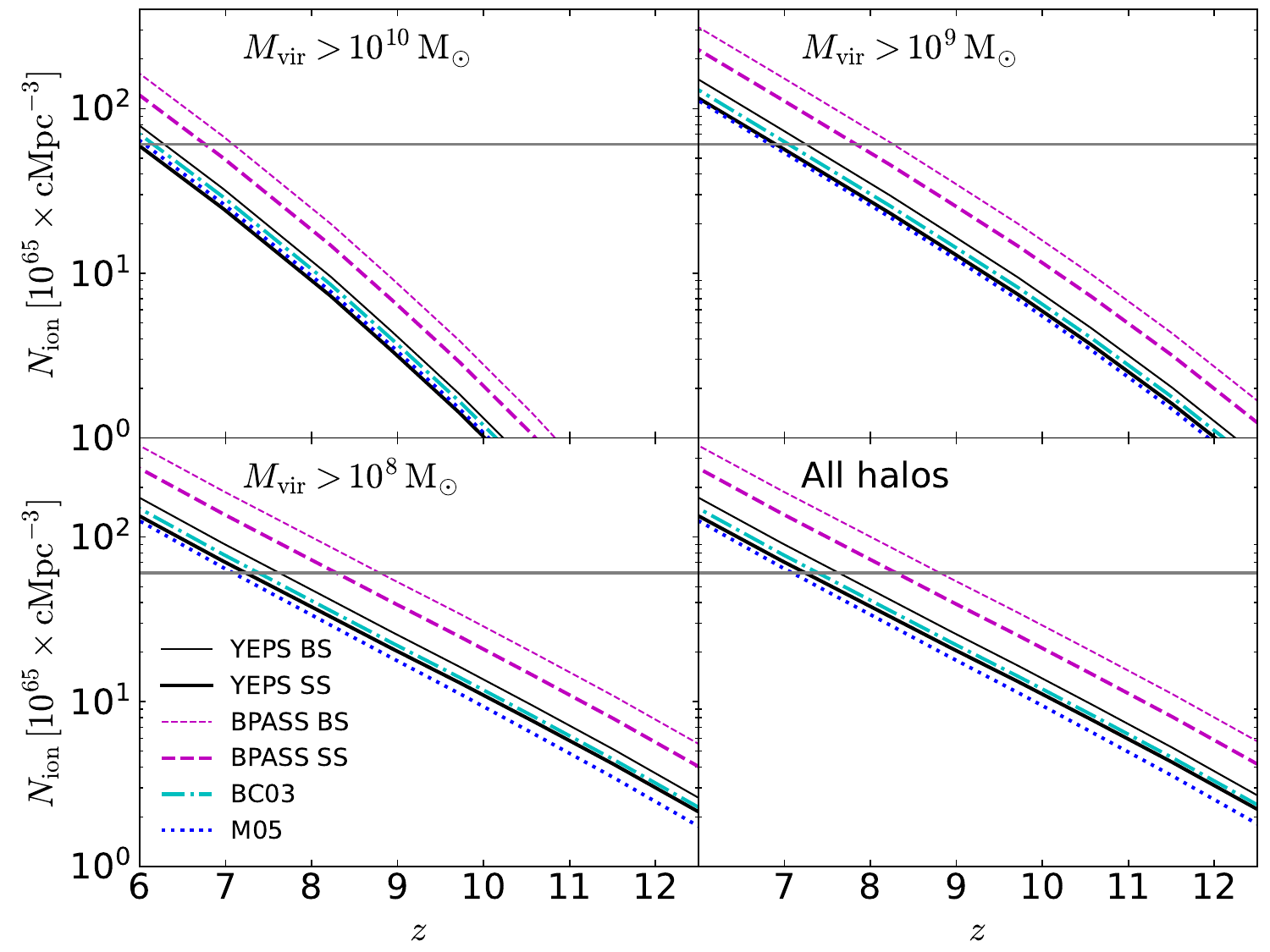}
    \caption{Redshift evolution of volume averaged ionizing photon $N_{\rm ion}$ from four SPS models i.e. BC03 (dash-dotted cyan), M05 (dotted blue), YEPS (solid black) and BPASS (dashed magenta).
    The thin lines of YEPS (YEPS~BS) and BPASS (BPASS~BS) models are the results including binary stars, while the thick lines (i.e. YEPS~SS and BPASS~SS) are with single stars.
    From left to right and top to bottom, the four panels are the $N_{\rm ion}$ of galaxies with $M_{\rm vir} > 10^{10}\,\rm M_{\odot}$, $> 10^{9}\,\rm M_{\odot}$, $> 10^{8}\,\rm M_{\odot}$ and all halos.
    As a reference, the horizontal gray line is the minimal budget of ionizing photon number density to fully ionize neutral hydrogen and first ionizing of helium}.
    \label{fig:ion_vz}
\end{figure}
Fig.~\ref{fig:ion_vz} shows the $N_{\rm ion}$ of galaxies within different halo mass $M_{\rm vir}$ range from four SPS models as functions of $z$s.
As a comparison, we also present the minimal budget of ionizing photon number density to fully ionize neutral hydrogen and first ionizing of helium by assuming 75\% of baryon is hydrogen and 25\% is helium, and show as the horizontal gray lines.
From Fig.~\ref{fig:ion_vz}, the ionizing photons are mostly from the galaxies with $M_{\rm vir} > 10^{9}\,\rm M_{\odot}$, and $\sim$ half of them are from the galaxies with $M_{\rm vir} > 10^{10}\,\rm M_{\odot}$.
The galaxies with $M_{\rm vir} < 10^{9}\,\rm M_{\odot}$ only slightly increase the $N_{\rm ion}$.
This is partly due to the incomplete sample of halos with $M_{\rm vir} < 10^{9}\,\rm M_{\odot}$ from MS-II simulation.
Meanwhile, the supernovae and radiation feedback also suppresses the star formation on such low mass halos \citep{Hutter2021, Legrand2023}.
In Fig.~\ref{fig:ion_vs_stellar} of Appendix~\ref{app:pdf_ion}, we show one sample of the distributions of ionizing photons at $z=7$ versus halo mass.

Since the different SPS models and the binary stars mostly affect the radiation of galaxies at high energy band ($h_{\rm p}\nu > 13.6 \,\rm eV$) as showed in Fig.~\ref{fig:output_sed_multi_z}, the four SPS models have much larger differences on the budget of ionizing photons (i.e. $N_{\rm ion}$) than that on the UV luminosity function showed in Fig.~\ref{fig:uvlf_multi_z}.
Specifically, the BPASS model has the highest $N_{\rm ion}$, which is $\sim 2$ times that of YEPS model.
Comparing with the minimal budget of ionizing photon number density to fully ionize neutral hydrogen and first ionizing of helium, this can lead to redshift difference $\delta z \sim 1$ on the end redshift of EoR.
Note that, the precise end redshift of EoR should be computed with the radiative transfer simulations, i.e. the redshift difference estimated might be different due to the ionizing and recombination models.
The M05 model has $N_{\rm ion}$ similar to that of YEPS model, which is slightly lower than that of BC03 model.
With the inclusion of binary stars, the BPASS model (BPASS~BS) can have $\sim 40\%$ more budget of $N_{\rm ion}$ than that with single stars (BPASS~SS), consistent with the results of e.g. \cite{Ma2022}, while the YEPS model (YEPS~BS) has $\sim 28\%$ higher $N_{\rm ion}$ than YEPS~SS.
The differences on $N_{\rm ion}$ caused by the binary stars are much smaller than that due to different SPS models.
Such results are consistent with the SEDs of galaxies showed in Fig.~\ref{fig:output_sed_multi_z}.

\begin{figure}
\centering
    \includegraphics[width=0.98\linewidth]{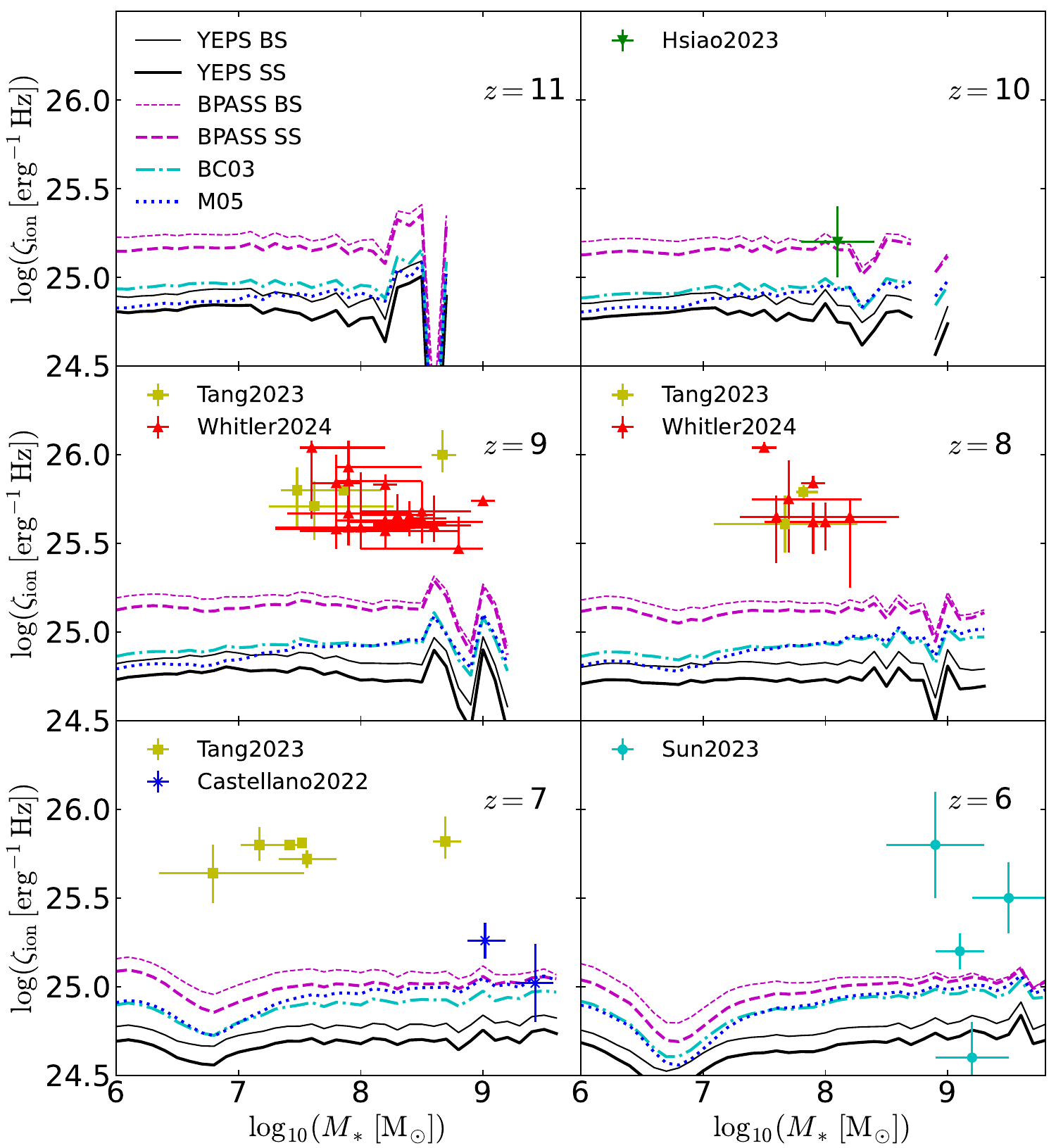}
    \caption{Ionizing photon production efficiency $\zeta_{\rm ion}$ from four SPS models i.e. BC03 (dash-dotted cyan), M05 (dotted blue),  YEPS (solid black) and BPASS (dashed magenta).
    The thin lines of YEPS (YEPS~BS) and BPASS (BPASS~BS) models are the results including binary stars, while the thick lines (i.e. YEPS~SS and BPASS~SS) are with single stars.
    From left to right and top to bottom, the six panels are the results at $z = 11$, 10, 9 8, 7 and 6, respectively.
    The observational data points are from \citealt{Hsiao2023} (green down triangle), \citealt{Whitler2024} (red up triangle),  \citealt{Tang2023} (yellow square), \citealt{Castellano2022} (blue star) and \citealt{Sun2023}  (cyan cycle).
    }
    \label{fig:ioneff_multi_z}
\end{figure}
To compare with the observations, in Fig.~\ref{fig:ioneff_multi_z} we show the ionizing photon production efficiency $\zeta_{\rm ion}$ from four SPS models, which is defined as:
\begin{equation}
    \zeta_{\rm ion} = \frac{\dot{n}_{\rm ion}}{F_{\rm 1600}}
\end{equation}
where $\dot{n}_{\rm ion}$ is the ionizing photon emissivity of galaxies.
Our results are roughly consistent with the previous studies e.g. \cite{Wilkins2016, Yung2020a, Seeyave2023}.
We only present some results of recent measurements about $\zeta_{\rm ion}$, while for more ones one can refer to the recent paper \cite{Simmonds2024}.
From Fig.~\ref{fig:ioneff_multi_z}, the $\zeta_{\rm ion}$ from our simulations are lower than the measurements by \cite{Tang2023, Whitler2024}, while roughly consistent with other observations.

%%%%%%%%%%%%%%%%%%%%%%%%%%%%%%%%%%%%%%%%%%%%%%%%%%
\section{Discussion and Conclusion}
\label{sec:conclusion}

The stellar population synthesis (SPS) model is an important component within galaxy formation models.
The uncertainties on the predictions of SPS models will affect the theoretical studies of galaxy formation and reionization process with the observations by the telescopes.
In this paper, we use the semi-analytical galaxy formation model {\sc L-Galaxies 2020} \citep{Henriques2020} and the {\it N}-body dark matter simulation Millennium-II \cite[MS-II, ][]{Boylan-Kolchin2009} to explore the effects of four popular SPS models, i.e. BC03 \citep{Bruzual2003}, M05 \citep{Maraston2005}, YEPS \citep{Zhang2005} and BPASS \citep{Stanway2018}, on the properties of high-$z$ galaxies and the budget of ionizing photons during epoch of reionization (EoR), including both the modes with binary stars and only single stars.

We find the uncertainties of different SPS models on the SEDs of galaxies are mostly at the high energy band, i.e. $h_{\rm P}\nu > 13.6\,\rm eV$.
With this reason, the four SPS models have not significant differences on the UV luminosity functions which are measured at the rest-frame wavelength $\lambda = 1600$~\AA, while predict obviously different budget of ionizing photon number density ($N_{\rm ion}$), consistent with the conclusions of e.g. \cite{Wilkins2016, Yung2020a}.
Specifically, the BPASS model has the higher amplitudes of SED than other models at $h_{\rm P}\nu < 50\,\rm eV$, which predicts $N_{\rm ion} \sim 2$ times that of YEPS model.
The BC03 and M05 models predict similar SEDs, and thus have similar $N_{\rm ion}$, which are slightly higher than the YEPS model but much lower than the BPASS model.
The inclusion of binary stars does not visibly change the UV luminosity functions, while predicts $\sim 40\%$ more ionizing photons with the BPASS model and $\sim 28\%$ more with the YEPS model.
We note that, the BPASS model adopts a detailed stellar evolution calculation with the Cambridge STARS code instead of the approximate and rapid stellar evolution model applied in e.g. the YEPS model \citep{Han2019}, which results should be more credible.

Considering that the differences of different SPS models on the high-$z$ galaxies are mostly on the ionizing band which can be highly absorbed by the neutral hydrogen, we do not expect the JWST observations can directly distinguish different SPS models. The indirect measurements e.g. the ionizing photon production efficiency $\zeta_{\rm ion}$ measured by the SED fitting might help to distinguish different SPS models \citep{Seeyave2023}.

In this paper, we only focus on the effects of using different SPS models.
All the initial SEDs from four SPS models are the public data from their official websites.
We take the ones with the same initial mass functions (IMF) i.e. Salpeter IMF.
We note that the different assumptions of IMF, stellar evolution, stellar atmosphere and/or binary star model will also change the predictions on the properties of high-$z$ galaxies and the budget of ionizing photons \citep{Seeyave2023}.
The specific modeling of each SPS and the detailed explanation for their difference are beyond the scope of this paper.

Although the {\sc L-Galaxies 2020} code includes most of the physical processes related to galaxy formation, while not the effects of radiation feedback, which can suppress the star formation on halos $<10^{9}\,\rm M_{\odot}$ \citep{Hutter2021, Legrand2023}.
However, we do not expect it will obviously change the results in this paper, as the star formation of halos $<10^{9}\,\rm M_{\odot}$ is already reduced by the supernovae feedback.
We adopt the best-fit parameters in the L-Galaxies 2020 model with the MS-II simulation and the low-$z$ observations, i.e. assuming that all the models at low-$z$ are still available for the high-$z$ ones.
However, we can expect that different dark matter simulations and SAM codes might lead to different results for high-$z$ galaxies.

As a summary, although the different SPS models and binary stars will not significantly change the high-$z$ UV luminosity functions, they obviously affect the prediction of the budget of ionizing photons during EoR.
Thus, the proper selection of SPS model is important to study the EoR process, especially when including the galaxy formation process in the modeling of EoR.

\begin{acknowledgments}
This work is supported by the National SKA Program of China (grant No. 2020SKA0110402), National Natural Science Foundation of China (Grant No. 12263002), and GZNU 2019 Special projects of training new academics and innovation exploration.
Yunkun Han gratefully acknowledges the support from the National Key R\&D Program of China (Nos. 2021YFA1600401 and 2021YFA1600400), the National Science Foundation of China (grant nos. 11773063, 12288102) the China Manned Space Project (grant nos. CMS-CSST-2021-A02, CMS-CSST-2021-A04, CMS-CSST-2021-A06), the 'Light of West China' Program of Chinese Academy of Sciences, the Yunnan Ten Thousand Talents Plan Young \& Elite Talents Project, the Natural Science Foundation of Yunnan Province (No. 202201BC070003), and the International Centre of Supernovae, Yunnan Key Laboratory (No. 202302AN360001).
The tools for bibliographic research are offered by the NASA Astrophysics Data Systems and by the JSTOR archive.
\end{acknowledgments}

\appendix

\section{One example of full SED from four SPS models}
\label{app:full_sed}
\begin{figure}
\centering
    \includegraphics[width=0.95\linewidth]{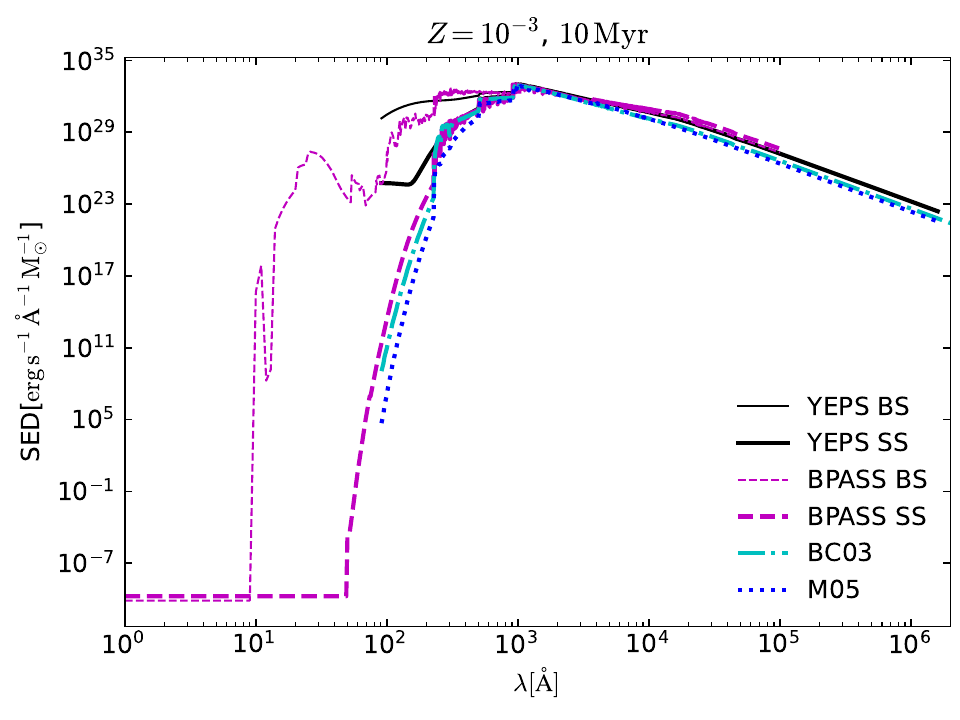}
    \caption{Stellar mass normalized SED from SPS model BC03 (dash-dotted cyan), M05 (dotted blue), YEPS (solid black) and BPASS (dashed magenta), with wavelength $\lambda$ as the x-axis.
    The results are at metallicity $Z=10^{-3}$ and age 10 Myr.
    The thin lines of YEPS (YEPS~BS) and BPASS (BPASS~BS) models are the SEDs including binary stars, while the thick lines (i.e. YEPS~SS and BPASS~SS) are with single stars.}
\label{fig:origizal_sed_inputs_full}
\end{figure}
Fig.~\ref{fig:origizal_sed_inputs_full} shows one example of full SED from four SPS models at metallicity $Z=10^{-3}$ and age 10 Myr.
Four SPS models have similar SEDs at $\lambda \gtrsim 900$~\AA, while show some differences at shorter $\lambda$.
The inclusion of binary stars obviously increases the amplitudes of SED at $\lambda \lesssim 900$~\AA, both with the YEPS and BPASS models.

\section{Distributions of ionizing photons at $z=7$ as functions of halo mass}
\label{app:pdf_ion}
\begin{figure}
\centering
    \includegraphics[width=0.98\linewidth]{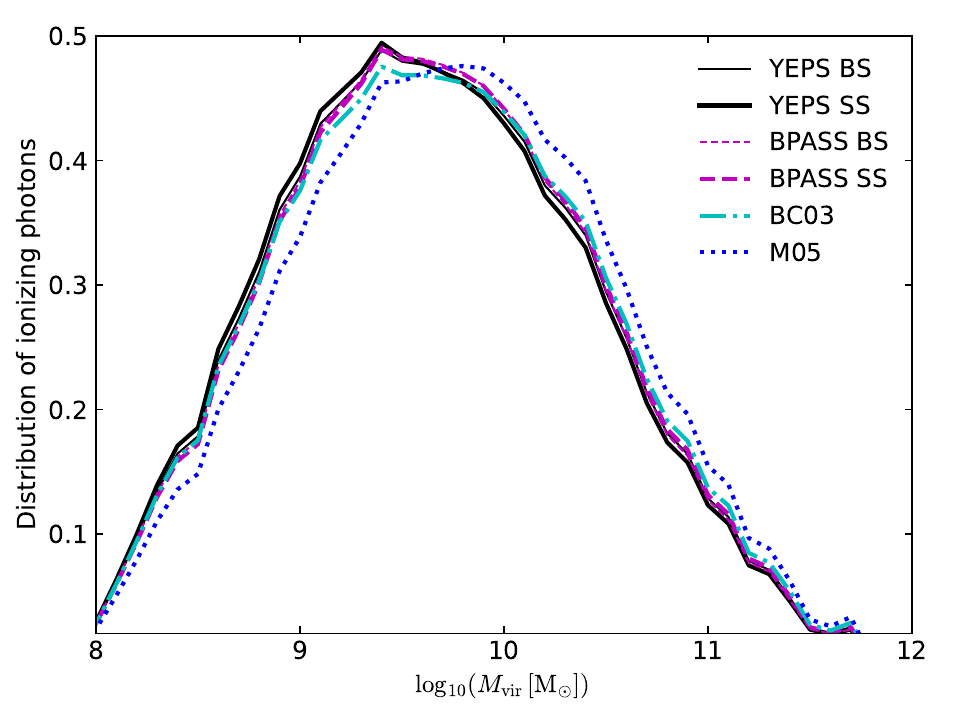}
    \caption{Distributions of ionizing photons at $z=7$ from SPS model BC03 (dash-dotted cyan), M05 (dotted blue), YEPS (solid black) and BPASS (dashed magenta) as functions of halo mass $M_{\rm vir}$.
    The thin lines of YEPS (YEPS~BS) and BPASS (BPASS~BS) models are the results including binary stars, while the thick lines (i.e. YEPS~SS and BPASS~SS) are with single stars.
    }
\label{fig:ion_vs_stellar}
\end{figure}
Fig.~\ref{fig:ion_vs_stellar} shows the distributions of ionizing photons at $z=7$ from four SPS models, which is computed by $\Delta (n_{\rm ion}) / \Delta({\rm log_{10}}(M_{\rm vir})) / \sum n_{\rm ion}$, where $\Delta (n_{\rm ion})$ is sum of $n_{\rm ion}$ from the halos within a mass bin width $\Delta({\rm log_{10}}(M_{\rm vir}))=0.1$, and $\sum n_{\rm ion}$ is the total number of ionizing photons from all halos.
From Fig.~\ref{fig:ion_vs_stellar}, the ionizing photon number increases with halo mass decreasing until $M_{\rm vir} \approx 3 \times 10^{9} \rm M_{\odot}$, while decreases with halo mass decreasing at lower $M_{\rm vir}$.
This conclusion is not very sensitive to the SPS models except the M05 one.

\bibliography{ref}{}
\bibliographystyle{aasjournal}



\end{document}